\documentclass[useAMS,usenatbib]{mn2e}

\usepackage{graphicx}
\usepackage{epsfig}
\usepackage{times}
\usepackage{natbib}
\usepackage{amsfonts}

\def\bm#1{{\hbox{\boldmath $#1$\unboldmath}}}

\newcommand{\be}{\begin{equation}}
\newcommand{\ee}{\end{equation}}
\newcommand{\bea}{\begin{eqnarray}}
\newcommand{\eea}{\end{eqnarray}}

\newcommand{\av}{{\rm{av}}}
\newcommand{\Max}{{\rm{max}}}
\newcommand{\g}{{\mathcal{G}}} 
\newcommand{\pol}{{{\rm pol}}}

\newcommand{\covT}{{C}}
\newcommand{\covTL}{{C_l^{\Delta T_{\rm obs}}}}
\newcommand{\cov}{{\tilde C}}
\newcommand{\covL}{{\tilde C(l)}}
\newcommand{\covTred}{{C_{\rm red}}}
\newcommand{\covTredInv}{{C_{\rm red}^{-1}}}
\newcommand{\covTredL}{{C_l^{\rm red}}}

\newcommand{\Ttempl}{{T_\tau}}
\newcommand{\Tdet}{{T_{\rm det}}}
\newcommand{\Tobs}{{T_{\rm obs}}}
\newcommand{\Tcmb}{{T_{\rm cmb}}}
\newcommand{\Tfg}{{T_{\rm fg}}}
\newcommand{\Ts}{{T_{\rm s}}}
\newcommand{\Tred}{{T_{\rm red}}}
\newcommand{\TerrObs}{{\Delta T_{\rm obs}}}
\newcommand{\Tisw}{{T_{\rm isw}}}
\newcommand{\Tprim}{{T_{\rm prim}}}

\newcommand{\Etempl}{{E_\tau}}
\newcommand{\Edet}{{E_{\rm det}}}
\newcommand{\Eobs}{{E_{\rm obs}}}
\newcommand{\Ecmb}{{E_{\rm cmb}}}
\newcommand{\Efg}{{E_{\rm fg}}}
\newcommand{\Es}{{E_{\rm s}}}
\newcommand{\ErrEobs}{{\Delta E_{\rm obs}}}

\title[Ironing out primordial temperature fluctuations with polarisation] 
{Ironing out primordial temperature fluctuations with polarisation:
 optimal detection of cosmic structure imprints} 
\author[M. Frommert \& T.~A. En{\ss}lin]
{M. Frommert \& T.~A. En{\ss}lin\\
Max-Planck-Institut f\"ur Astrophysik,
Karl-Schwarzschild-Stra{\ss}e 1, D-85748 Garching b. M\"unchen, Germany\\
mona@mpa-garching.mpg.de\\
}

\begin{document}

\date{Accepted ??? Received ???; in
  original form ???}
\pagerange{\pageref{firstpage}--\pageref{lastpage}} \pubyear{2008}
\maketitle
\label{firstpage}

\begin{abstract}
Secondary anisotropies of the cosmic microwave background (CMB) can be
detected by  
using the cross-correlation between the large-scale structure (LSS)
and the CMB temperature fluctuations. 
In such studies, chance correlations of primordial 
CMB fluctuations with the LSS are the main source of uncertainty.
We present a method for reducing this noise by exploiting information
contained in the polarisation of CMB
photons. The method is described in general terms and then applied to
our recently proposed optimal method for measuring the integrated
Sachs-Wolfe (ISW) effect.
We obtain an expected signal-to-noise ratio of up to 8.5.
This corresponds to an enhancement of the signal-to-noise by 23 per
cent as compared to the standard method for ISW detection, and by 16 per cent
w.r.t. our recently proposed method, both for the best-case scenario
of having perfect (noiseless) CMB and LSS data.

\end{abstract}



\section{Introduction} \label{intro}

The low-redshift large-scale structure (LSS) changes the cosmic
microwave background (CMB) fluctuations in various ways. Such
secondary effects on the CMB are, for example, the integrated
Sachs-Wolfe (ISW) effect \citep{isw}, the Rees-Sciama (RS) effect
\citep{rs}, gravitational lensing \citep{lensing}, and the
Sunyaev-Zel'dovich (SZ) effect \citep{sz_1972, sz_1980}.
By studying these signals, we can obtain valuable information about
our Universe. The ISW effect,
for example, provides independent evidence for the existence of dark
energy. 
Unfortunately, unless the spectral
signatures of the signal differ from the ones of the primordial CMB,
it is difficult to detect them. The reason is that 
the primordial CMB fluctuations created at the time of last scattering
are much stronger than the secondary temperature anisotropies.
The usual method for detecting secondary anisotropies in the CMB
is via cross-correlating the CMB temperature maps with LSS
data such as the galaxy density contrast. Since secondary
anisotropies in the CMB are created by the LSS, there is a
significant cross-correlation between the two. In contrast, the primordial CMB
fluctuations should not be
correlated with the LSS. By performing the
cross-correlation analysis, one can therefore separate signatures of
the presence 
of these secondary anisotropies from the primordial fluctuations.

In the standard cross-correlation method, first described by
\cite{x_ray_boughn}, the observed cross-correlation between LSS
and CMB data is compared to its theoretical prediction. This method has been
extensively used to detect the ISW effect. Some of the most recent
studies are by \cite{ho}, \cite{giannantonio},
\cite{rassat}, and \cite{boughn_crittenden_nature}.
Since the theoretical cross-correlation function is by
construction an ensemble average over all possible
universes, fluctuations associated with the specific
realisation of the LSS in the observed Universe
act as a source of noise in the detected signal in the standard method. 

In \cite{mona}, we suggested a method for reducing this source of
uncertainty, which we will refer to as the optimal temperature-only method.
Instead of comparing the
observed cross-correlation function with its theoretical prediction,
we use an optimal matched filter in order to detect an ISW template in
CMB data.
Similar schemes were independently
proposed by \cite{zhang}, \cite{carlos} and \cite{szapudi}. 
Optimal matched filters have also been used to study other secondary
effects on the CMB. The first of these studies explored the detectability of
the kinetic SZ effect of galaxy clusters \citep{haehnelt_and_tegmark},
later works on the kinetic SZ effect and the RS effect are for example
by \cite{bjoern}, \cite{matteo_rs}, \cite{matteo_rs_filter}, and \cite{andre}.

However, in both the standard method and the optimal temperature-only
method, the main source of uncertainty in the detection of the
secondary signal comes 
from chance correlations of primordial CMB fluctuations
with the LSS.
In this work, we present a method which exploits polarisation
information in order to reduce
not only the noise from the specific LSS realisation, but 
also the noise coming from primordial CMB temperature fluctuations.
This method can be
applied generically to the detection of all secondary effects. It is
based on the fact that the polarisation measured in the CMB contains
information about the primordial temperature fluctuations. 
We use the observed E-mode polarisation map, which we translate into a
temperature map using the
TE cross-power spectrum. The obtained temperature map is then
subtracted from the observed temperature map, and hence
no longer contributes to the noise budget of the detected
signal. Once an E-map has been measured to a good
accuracy, this will significantly 
enhance the signal-to-noise ratio of the detection of secondary effects.
The first all-sky measurement of polarisation with 
high fidelity is expected to be provided
by the Planck Surveyor satellite \citep{planck}, to be launched in 2009.

Our optimal polarisation method builds on the optimal scheme to
detect LSS signatures in CMB data, which we developed in 
\cite{mona} specifically for the ISW effect. Note that this method
assumes a Gaussian data model, hence it is very well suited for the
ISW effect, whereas one might need to extend it into the non-Gaussian
regime for other effects, such as the RS effect, the kinetic SZ effect
or lensing. This can be 
done using information field theory \citep{ift}, but
is beyond the scope of this work. Here we show how to use the information
contained in polarisation data within the framework of a Gaussian
data model and leave the extension to more complicated models for
future work.

When applying our method to ISW detection, we obtain an expected
signal-to-noise ratio of up to 8.5. This corresponds to an enhancement
of the signal-to-noise ratio by 16 per cent w.r.t. the optimal
temperature-only method, independent of the depth of the
galaxy survey considered. In comparison to the standard
method, the signal-to-noise ratio is enhanced by 23 per cent for a
full-sky LSS survey that goes out to redshift 2. Both of these
comparisons have been made for the best-case scenario of having
perfect (noiseless) CMB and LSS data. 

Using polarisation data to reduce the noise in the detection of
 secondary effects was first proposed by Robert Crittenden, following a
suggestion from Lyman Page \citep{crittenden}. He already derived the
reduced temperature power spectrum, which we show in Figure 
\ref{spectra}, and roughly estimates the improvement of the
signal-to-noise ratio for ISW detction to be around 20 per cent,
which we confirm with our calculations.

Our article is organised as follows. In section \ref{optimal_method}
we describe the optimal method derived in \cite{mona} in general terms. In
section \ref{pol} we then show how we can reduce the noise coming from 
primordial temperature fluctuations by using polarisation data. In section
\ref{isw}, we apply the method to the ISW effect. We conclude in section
\ref{conclusions}.

\section{Optimal method for the detection of secondary effects on the
  CMB} \label{optimal_method} 

In \cite{mona}, we derived an optimal method for the detection of
secondary temperature anisotropies in the CMB using as
example the ISW effect. In this section we briefly review this method, which
we refer to as the optimal temperature-only method.

Let's assume that we know the LSS well enough to create a template $\Ttempl$ of
the secondary signal $\Ts$ that we would like to detect in the temperature
fluctuations, for example the ISW signal
$\Tisw$. Here, $T_X$ with any index $X$ denotes the function $T_X: S^2
\rightarrow \mathbb{R}$, which we regard also as an element of a
function-vector space.
The data $d$ we measure are the observed CMB temperature
fluctuations $\Tobs$. Our data model is then
\bea \nonumber
d &\equiv& \Tobs \\ \nonumber
&=& \Tcmb + \Tfg + \Tdet\\ \nonumber
&=& \Ttempl + (\Tcmb - \Ttempl) + \Tfg + \Tdet \\ 
&\equiv& \Ttempl + \TerrObs,
\eea
where $\Tcmb$ denotes the cosmological CMB temperature fluctuations, $\Tfg$
are residual galactic foregrounds after foreground-removal,
and $\Tdet$ denotes the 
detector noise. Note that $(\Tcmb - \Ttempl) = (\Tcmb - \Ts) + (\Ts - \Ttempl)$
contains the the CMB fluctuations other than the signal we are after,
$(\Tcmb - \Ts)$,
and the uncertainty in the template w.r.t. the signal, $(\Ts -
\Ttempl)$, coming from 
our ignorance of the full distribution of the matter in the Universe.
Note that for simplifying the notation, we have redefined $T
\equiv \left(T - T_0\right)/T_0$, where $T_0$ denotes the monopole of
the CMB temperature fluctuations. 
An overview over the above definitions can be found in table~\ref{tab:tabelle}.
\begin{table}
\begin{tabular}{ l l }
  Symbol & Definition \\
\hline
  $\Tcmb$, $\Ecmb$ & cosmological CMB temperature and polarisation \\
  $\Ts$, $\Es$ & real secondary signal that we are trying to detect\\
  $\Ttempl$, $\Etempl$ & signal templates for temperature and polarisation \\
  $\Tfg$, $\Efg$ & residual galactic foregrounds after foreground removal\\
   $\Tdet$, $\Edet$ & detector noise \\
  $\Tobs$, $\Eobs$ & $(\Tcmb + \Tfg + \Tdet)$, $(\Ecmb + \Efg + \Edet)$ \\
  $\TerrObs$, $\ErrEobs$ & $(\Tobs - \Ttempl)$, $(\Eobs - \Etempl)$ \\
$\Tisw$ & fluctuations created by ISW effect \\
$\Tprim$ & $(\Tcmb - \Tisw)$ \\
\end{tabular}
\caption{Summary of defined symbols}
 \label{tab:tabelle}
\end{table}

We now approximate the distribution of $\TerrObs$ by a Gaussian around
zero. That
is, we write the
probability density function of $\Tobs$ given the signal template
$\Ttempl$ and the cosmological parameters $p$, the likelihood, as
\be \label{posterior}
P(\Tobs \,|\, \Ttempl , p) = \g\left(\Tobs - \Ttempl , \covT\right).
\ee
Here we have defined 
\be 
\g(\chi,C) \equiv \frac{1}{\sqrt{|2\pi C|}}
\exp \left(-\frac{1}{2}\chi^\dagger\,C^{-1}\chi \right)
\ee
to denote the probability density function 
of a Gaussian distributed vector $\chi$ with zero mean, given
the cosmological parameters $p$ and the covariance matrix $C \equiv \langle
\chi \chi^\dagger \rangle$, where the averages are taken over the
Gaussian distribution $\g(\chi ,C)$. Note that in general the covariance
matrix depends on the cosmological parameters, which is not
explicitly stated in our notation.
A daggered vector or matrix denotes its transposed and complex
conjugated version, as usual. 
Hence, given two vectors $a$ and $b$, $a\, b^\dagger$ must be
read as the tensor product,
whereas $a^\dagger \,b$ denotes the scalar product. 
Note that in
eq. (\ref{posterior}) the signal template $\Ttempl$ may depend on the
cosmological parameters $p$ as well.

Let us briefly address the question of how to create the
template $\Ttempl$. When writing down the likelihood in eq. (\ref{posterior}),
we have implicitely assumed that the template $\Ttempl$ is the mean of $\Tobs$
w.r.t. the probability distribution given in
eq. (\ref{posterior}). This probability distribution is conditional on
the template $\Ttempl$, or, in other words, conditional on the LSS
data $\delta_g$, from which we have created our template according to some
prescription. Note that usually $\delta_g$ denotes the galaxy density
contrast, but we use it to denote the LSS data in a more general sense
here, which could also be lensing information, for example.

In the following, we assume that
the signal $\Ts= R \,\delta_m$ is given by a linear operator $R$ applied to the
matter density contrast $\delta_m$. For the ISW
effect, the operator $R$ is explicitely derived in \cite{mona}.
We can then write
\bea \nonumber
\Ttempl 
&\equiv& \langle \Tobs \rangle_{P(\Tobs \,|\, \delta_g,p)} \\ \nonumber
&\approx& \langle \Ts \rangle_{P(\Ts \,|\, \delta_g,p)} + \langle (\Tcmb - \Ts)
\rangle_{P((\Tcmb - \Ts) \,|\, \delta_g,p)} \\ \nonumber 
&&+ \langle \Tfg
\rangle_{P(\Tfg \,|\, \delta_g,p)}+ \langle \Tdet \rangle_{P(\Tdet)}
\\
&=& R \langle \delta_m \rangle_{P(\delta_m \,|\, \delta_g,p)}, 
\label{wiener}
\eea
where we have used that $\Ts$, $(\Tcmb - \Ts)$, $\Tfg$ and
$\Tdet$ are approximately stochastically independent in the first
step, and that the three errors have vanishing means, 
$\langle  (\Tcmb -
\Ts) \rangle_{P((\Tcmb - \Ts) \,|\, \delta_g , p)} = \langle \Tfg
\rangle_{P(\Tfg \,|\, 
  \delta_g,p)} = \langle  \Tdet \rangle_{P(\Tdet)} =  0$, in the second
step. 
For the ISW effect, $(\Tcmb - \Ts) \equiv
(\Tcmb - \Tisw) = \Tprim$ are simply the primordial fluctuations,
which do have zero mean \citep{mona}. For other 
secondary effects, $\langle  (\Tcmb -
\Ts) \rangle_{P((\Tcmb - \Ts) \,|\, \delta_g , p)} = 0$ is probably still
a reasonably good approximation.  
In the last step, we have pulled the operator $R$ out of the mean.

We see that for creating the signal template $\Ttempl$,  we need the mean of the
matter density contrast conditional 
on the LSS data, $\langle
\delta_m \rangle_{P(\delta_m \,|\, \delta_g,p)}$. In the simplest case
of having a
Gaussian likelihood and Gaussian prior for $\delta_m$, this is given
by the Wiener filter. Again this is a very good approximation for the
ISW effect, which is present on very large scales, on which
structure growth is still linear.
For other
effects such as the kinetic SZ effect, the RS effect or lensing, the
Gaussian approximation 
for $\delta_m$ may not be very good (thus also the Gaussian
approximation for $\TerrObs$ may not be good), and one would have to
consider non-Gaussian data models using information field theory
\citep{ift}. However, in this work we will use the Gaussian data model
and leave extensions to non-Gaussian models for future work.
Note that, when choosing the template as in eq. (\ref{wiener}), the
the latter is uncorrelated with $\TerrObs \equiv (\Tobs - \Ttempl)$
(w.r.t. the probability distribution in eq. (\ref{posterior})),
as can be easily shown.

In order to see how well we can recover such a signal template from
the CMB data,
we put an amplitude $A_\tau$ in front of the signal $\Ttempl$ in
eq. (\ref{posterior}), and try to estimate its value from the data 
(the true value of this amplitude is one, of course):
\be
P(\Tobs \,|\, A_\tau, \Ttempl , p) = \g(\Tobs - A_\tau \Ttempl , \covT).
\ee
The maximum likelihood estimator $\widehat A_\tau$ for the amplitude
$A_\tau$ is
\be
\widehat A_\tau = \frac{\Tobs^\dagger \covT^{-1} \Ttempl}{\Ttempl^\dagger
  \covT^{-1} \Ttempl} = \frac{\sum_l (2l+1) \frac{\widehat
    C_l^{\,\Ttempl,\Tobs}}{\covTL}}{\sum_l (2l+1) \frac{\widehat
    C_l^{\,\Ttempl}}{\covTL}}. 
\ee
In the second equality, we have assumed that the knowledge of the
secondary anisotropy template is equally good in any direction, so
that the template uncertainty matrix is isotropic and fully described
by its spherical harmonics power spectrum. We will use this assumption
also in the following.
This permits us to evaluate the expressions in spherical harmonics space in
the second step. We have used the following definitions of the
power spectra and their estimators (we use a hat to denote estimators)
\bea
C_l^{X,Y} &\equiv& \langle a_{lm}^X a_{lm}^{Y\,*} \rangle, \\
C_l^X &\equiv& C_l^{X,X},\\
\widehat C_l^{X,Y} &\equiv& \frac{1}{2l+1} \sum_l Re \left( a_{lm}^X
a_{lm}^{Y\,*} \right),\\
\widehat C_l^X &\equiv&  \widehat C_l^{X,X},
\eea
where the $a_{lm}$ are defined by an expansion into spherical
harmonics $Y_{lm}$:
\be
a_{lm}^X \equiv \int_S d\Omega \, T_X(\bm{\hat n}) Y_{lm}^*(\bm{\hat n}). 
\ee
The power spectrum $\covTL$ denotes the spherical harmonics space
version of the covariance matrix $\covT$. 
We calculate the variance of the amplitude estimator to be
\bea \nonumber
\sigma_A^2 &\equiv& \langle \left( \widehat A_\tau - \langle \widehat A_\tau
\rangle_{\rm cond} \right)^2 \rangle_{\rm cond}\\ 
&=& \left( \Ttempl^\dagger
\covT^{-1}\Ttempl \right)^{-1} = \left( \sum_l(2l+1) \frac{\widehat
    C_l^{\,\Ttempl}}{\covTL}  \right)^{-1},
\eea
where we have again evaluated the expressions in spherical harmonics
space in the last step, and
we have used the notation introduced in \cite{mona}, where the
index ``cond'' indicates that the average is taken conditional on the
signal template $\Ttempl$, i.e. over the probability distribution given
in eq. (\ref{posterior}). 
We can now define the signal-to-noise ratio as follows
\be \label{StoNtau}
\left( \frac{S}{N} \right)_t^2 \equiv \frac{1}{\sigma_A^2} =
\sum_l (2l+1) \frac{ \, \widehat C_l^{\,\Ttempl}}{\covTL},
\ee
where the index $t$ indicates that this is the signal-to-noise ratio one
obtains for the optimal temperature-only method.
This signal-to-noise ratio depends on the actual realisation of the
matter distribution in our Universe via the estimator $\widehat
C_l^{\,\Ttempl}$. In \cite{mona}, we showed that for the ISW effect
we obtain on average a signal-to-noise ratio of about 7, if we
assume an ideal LSS survey which covers the whole sky and goes out to
a redshift of about 2. In comparison to the standard method, this is
an enhancement of the signal-to-noise ratio by about 7 per cent.

\section{Reduction of the primordial noise using polarisation
  information}\label{pol} 

With the method suggested in \cite{mona}, we were able to reduce the low
redshift cosmic variance effect in amplitude estimates of secondary signals,
i.e. we reduced the noise coming from the specific realisation of 
LSS in our Universe. Now we 
show how even the noise coming from primordial temperature
fluctuations can be reduced. The idea is that since the
temperature and polarisation maps of the CMB are correlated, the
polarisation contains information about the temperature fluctuations. After
extracting this information from the polarisation data we know a part
of the temperature map, which we can remove from the data before trying to
detect the signal. In other words, we make our amplitude estimate of
the secondary signal conditional on the known part of the
temperature fluctuations. 

To include the information contained in the polarisation data, we
enlarge our data vector $d$ to include the observed E-mode polarisation
map $\Eobs$ as well: 
\be
d \equiv \left(\Tobs,\Eobs\right)^\dagger,
\ee
or, in spherical harmonics space
\be
a_{lm}^d \equiv \left( a_{lm}^\Tobs, a_{lm}^\Eobs \right)^\dagger.
\ee
Note that with {\it the map} $\Eobs$, we are referring again to
the abstract element of a function-vector space space, which contains all the
information on the observed E-mode. When evaluating the abstract
expressions obtained in the following, we use the representation of $\Eobs$ in
spherical harmonics space, consisting of all coefficients $a_{lm}^\Eobs$.

In principle, it is possible that the secondary effect we are looking
for is also present as a small signal in the polarisation data. If the
temperature anisotropies created by the secondary effect exhibit
a quadrupole component at the time of reionization, this
quadrupole will be rescattered by free electrons and create a
polarisation signal \citep{zaldarriaga}. However, for the
ISW this effect has been proven to be small
\citep{cooray_pol}. It should also be small for the RS effect, lensing and
the kinetic SZ effect, the highest contributions of which are on relatively
small scales. 
Thus, as a first approximation we assume that the polarisation data do not
carry any signal of the effect we want to detect.
Our signal template $\tau$ is then
\bea \nonumber
\tau &\equiv& \left(\Ttempl,0\right)^\dagger, \\
a_{lm}^\tau &\equiv& \left(a_{lm}^\Ttempl,0\right)^\dagger,
\eea
and the data model becomes
\be
d = \left( \begin{array}{l}
\Tobs \\ \Eobs 
\end{array} \right) 
=
\left( \begin{array}{l}
\Ttempl + \TerrObs \\ \Eobs
\end{array} \right)\,.
\ee
The observed E-map, $\Eobs = \Ecmb + \Efg + \Edet$, consists of the
cosmological E-mode fluctuations $\Ecmb$, residual galactic foregrounds after
foreground removal $\Efg$, and the detector noise $\Edet$.
Assuming again Gaussianity, we can write down the likelihood
\be \label{post_pol}
P(d \,|\, \tau,p) = \g(d - \tau, \cov),
\ee
where the covariance matrix $\cov$ is
\be
\cov \equiv \langle (d-\tau)(d-\tau)^\dagger \rangle_{\rm cond},
\ee
and we have redefined the index 'cond' to denote the average
over the probability distribution in eq. (\ref{post_pol}). In
spherical harmonics
space, the covariance matrix $\cov$ is block-diagonal with the blocks being
\be \label{covdef}
\covL = \left( 
\begin{array}{cc}
\covTL & C_l^{\TerrObs,\Eobs} \\
C_l^{\TerrObs,\Eobs} & C_l^\Eobs  \\
\end{array} 
\right).
\ee
Therefore, the likelihood factorises:
\be \label{fact1}
P(d \,|\, \tau,p) = \prod_{l,m} \g(a_{lm}^d - a_{lm}^\tau , \covL).
\ee
When inserting the inverse of the covariance matrix $\covL$, it is possible to
rewrite the likelihood as a product of a reduced temperature
part and a polarisation part. To this end, let us define the reduced
temperature map and power spectrum
\bea \nonumber \label{red}
a_{lm}^\Tred &\equiv& a_{lm}^\Tobs -
\frac{C_l^{\TerrObs,\Eobs}}{C_l^\Eobs}a_{lm}^\Eobs,\\ 
\covTredL &\equiv& \covTL -
\frac{\left(C_l^{\TerrObs,\Eobs}\right)^2}{C_l^\Eobs}. 
\eea
With these definitions, the likelihood becomes
\be \label{fact}
P(d \,|\, \tau, p) = \prod_{l,m} \left[ \g(a_{lm}^\Tred - a_{lm}^\Ttempl,
  \covTredL)\,  \g(a_{lm}^\Eobs, C_l^\Eobs)  \right],
\ee
as we prove in Appendix \ref{prob_fact}. Now our goal is to find the signal
template $\Ttempl$ in the CMB data. The polarisation part of the above
likelihood, $\g(a_{lm}^\Eobs, C_l^\Eobs)$, does not
depend on the signal template, nor does the reduced temperature part
explicitely depend on $\Eobs$. In
other words, the observed E-map does not contain relevant information
any more after introducing the reduced temperature fluctuations.
Thus, we can marginalize over it, and continue
only with the likelihood of the reduced temperature map
\bea \nonumber
P(\Tred \,|\, \Ttempl, p) &\equiv& \g(\Tred - \Ttempl, \covTred) \\
&=& \prod_{l,m} \g(a_{lm}^\Tred - a_{lm}^\Ttempl,
\covTredL). 
\label{post_red}
\eea

Note that it is straightforward to derive the factorised likelihood
also for the case that we do have a non-zero signal template $\Etempl$ for the
polarisation part. In that
case, the covariance matrix $\covL$ is slightly changed, as well as
the definitions of the reduced temperature map and power
spectrum, and we can no longer neglect the polarisation part of the
likelihood. Please refer to Appendix \ref{prob_fact} for details.

Let us pause for a second and have a closer look at the quantities defined in
eq. (\ref{red}). What we have effectively done is the following. We have 
 a polarisation map $a_{lm}^\Eobs$, which is correlated with the temperature
fluctuations $a_{lm}^\TerrObs$ via $C_l^{\TerrObs,\Eobs}$. That is, the
polarisation map contains information about the temperature map, which we
can translate into a 'known' part of the temperature map using the
prescription $\left(C_l^{\TerrObs,\Eobs} / C_l^\Eobs\right) a_{lm}^\Eobs$.
This known part of the temperature map is subtracted from the observed
one, and we work only with the remaining unknown temperature
fluctuations in which we try to detect our signal template. 

The reduced temperature map fluctuates around our
signal template $\Ttempl$ only with the 
variance $\covTredL$, which is smaller than the full variance
$C_l^\TerrObs$ of the observed temperature map.
This reduced variance is
the uncertainty going into our signal detection problem now, rather
than the full variance of the original temperature fluctuations. 

In order to see this, let us again put an amplitude in front of the
signal template in 
eq. (\ref{post_red}), and estimate it from the data using a
maximum likelihood estimator:
\be
\widehat A_\tau = \frac{\Tred^\dagger \covTredInv \Ttempl}{\Ttempl^\dagger
  \covTredInv 
  \Ttempl} = \frac{ \sum_l (2l+1) \frac{\widehat
    C_l^{\Tred,\Ttempl}}{\covTredL} }{ 
\sum_l (2l+1) \frac{\widehat C_l^{\Ttempl}}{\covTredL} }.
\ee
Here, the last expression is in spherical harmonics space. The
variance of $\widehat A_\tau$ is now
\be
\sigma_A^2 = \left( \Ttempl^\dagger \covTredInv \Ttempl \right)^{-1} =
\left(\sum_l (2l+1) \frac{\widehat C_l^{\Ttempl}}{\covTredL} \right)^{-1}, 
\ee
and hence the signal-to-noise ratio becomes
\bea \nonumber
\left(\frac{S}{N}\right)_\pol^2 &=& \sum_l (2l+1) \frac{\widehat
    C_l^{\Ttempl}}{\covTredL} \\
&=& \sum_l \frac{(2l+1) \,\widehat C_l^{\Ttempl}}{\covTL -
\left(C_l^{\TerrObs,\Eobs}\right)^2 / C_l^\Eobs}.
\label{StoNpol}
\eea
Note that we have added the index ``pol'' to indicate that this is the
signal-to-noise ratio one obtains when using the polarisation data to
reduce the variance. 
Comparing the signal-to-noise ratio in
eq. (\ref{StoNpol}) with the one in eq. (\ref{StoNtau}), 
we see that by including the information contained in the polarisation
data, we reduce the variance in every mode by the term
$\left(C_l^{\TerrObs,\Eobs}\right)^2 / C_l^\Eobs$. 

\begin{figure}
 \centering
 \includegraphics{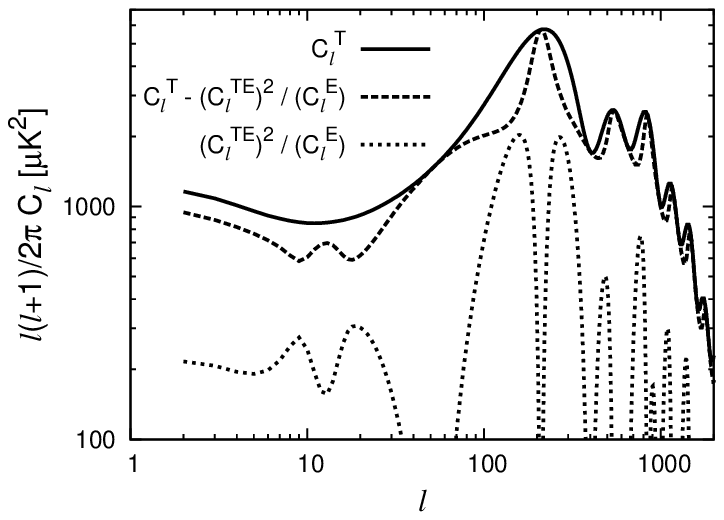}
 \caption{Reduction of the variance in the detection of secondary
   temperature signals by using the information contained
   in polarisation data.
   Shown are the CMB temperature power spectrum $C_l^\Tcmb$
   (solid), and the template-free part of
   the reduced temperature power spectrum $C_l^\Tcmb -
   \left(C_l^{\Tcmb,\Ecmb} \right)^2 / C_l^\Ecmb$ (dashed),
   together with the part of the CMB power spectrum coming from the
   'known' part of the temperature fluctuations which we infer from
   the polarisation map,
   $\left(C_l^{\Tcmb,\Ecmb} \right)^2 / C_l^\Ecmb$ (dotted).}
 \label{spectra}
\end{figure}
Let us now get an impression of how much the variance gets reduced for
the different multipoles. To this end, we neglect the detector noise $\Tdet$
and $\Edet$, and the foreground noise $\Tfg$ and 
$\Efg$\footnote{In reality, galactic E-mode foregrounds $\Efg$ are
  likely to be the 
  limiting factor in the improvement of the detection significance coming from
  including polarisation data. We comment
on this at the end of this section.}, which allows us write 
\bea \nonumber
C_l^{\TerrObs,\Eobs} &\approx& C_l^{\Tcmb,\Ecmb} - C_l^{\Ttempl,\Ecmb} \\
\covTL &\approx& C_l^\Tcmb - 2 C_l^{\Tcmb,\Ttempl} + C_l^\Ttempl\\ 
C_l^\Eobs &\approx& C_l^\Ecmb.
\eea
We furthermore neglect the cross-term
$C_l^{\Ttempl,\Ecmb}$. For the ISW effect,
we have verified numerically that it is negligible.
For the kinetic SZ and RS effects, the template itself is so
small that we can also certainly neglect $C_l^{\Ttempl,\Ecmb}$. 
Then, the reduced temperature power spectrum defined in eq. (\ref{red}) becomes
\be
\covTredL \approx C_l^\Tcmb - 2 C_l^{\Tcmb,\Ttempl} + C_l^\Ttempl -
\frac{\left(C_l^{\Tcmb,\Ecmb} \right)^2}{C_l^\Ecmb}.
\ee
In Fig. \ref{spectra}, we plot the template-free part of the reduced
temperature power spectrum $C_l^\Tcmb -
\left(C_l^{\Tcmb,\Ecmb} \right)^2 / C_l^\Ecmb$ (note that we have not
included the template-dependent terms $- 2 C_l^{\Tcmb,\Ttempl}$ and
$C_l^\Ttempl$ in the plot), which gives 
us an impression of how the variance coming from
primordial temperature fluctuations is being reduced by including
polarisation data. The variance will be further reduced by working
conditional on the signal template $\Ttempl$, which is encoded in the
terms $- 2 C_l^{\Tcmb,\Ttempl}$ and $C_l^\Ttempl$, and already described in
\cite{mona}. We also plot the original CMB power spectrum
$C_l^\Tcmb$ and the difference to the reduced one for comparison. We
have assumed a flat $\Lambda$CDM model with the parameter values given  
by \cite{wmap_5}, table 1 ($\Omega_b h^2 =
0.02265, \Omega_\Lambda = 0.721,\, h = 0.701,\, n_s = 0.96,\, \tau = 0.084,\,
\sigma_8 = 0.817$), and used CMBEASY
\citep[\texttt{www.cmbeasy.org,}][]{cmbeasy} for obtaining the  
respective spectra.

\begin{figure}
 \centering
 \includegraphics[trim = 15mm 0mm 15mm 0mm, clip, scale = 0.3, angle = 90]{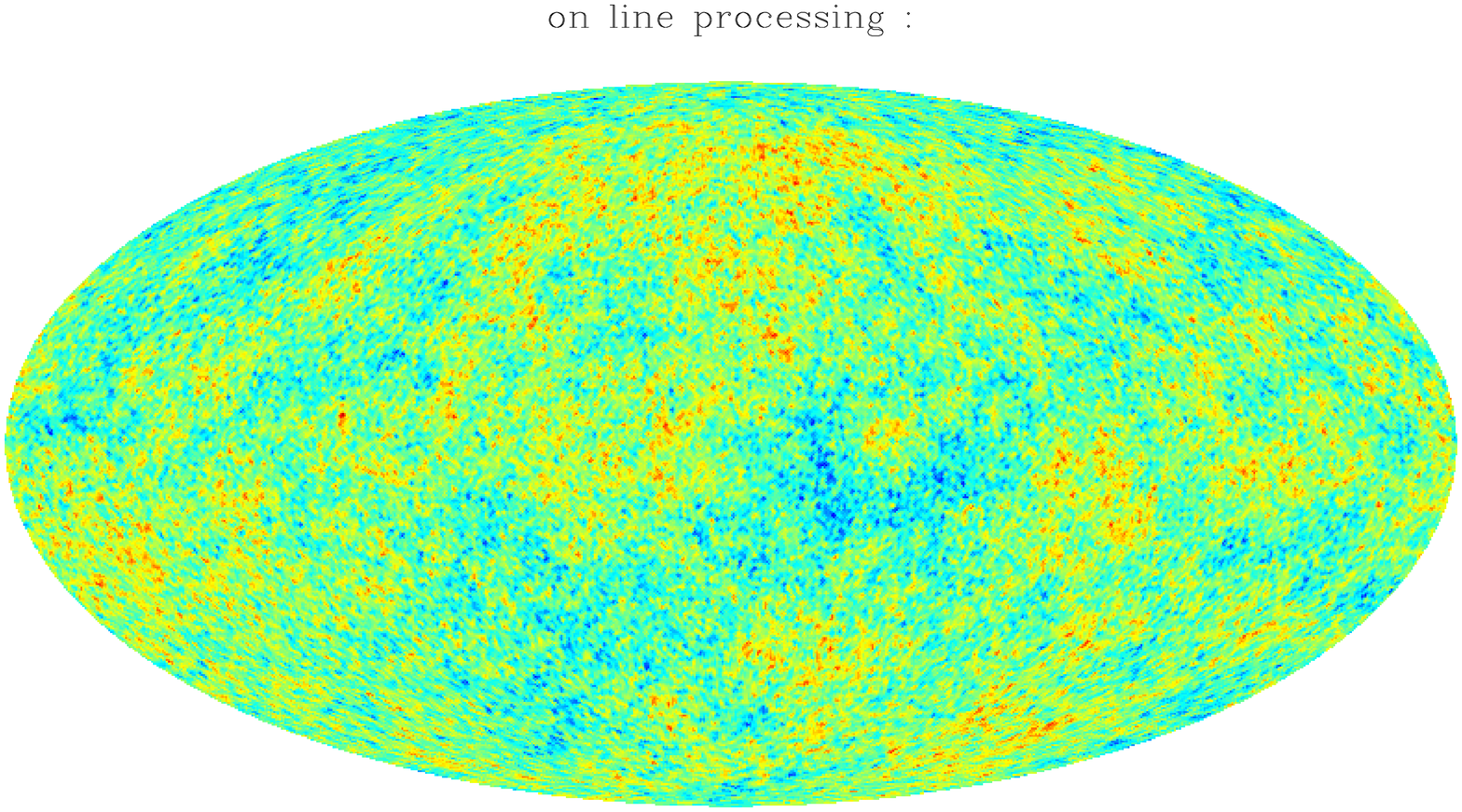}
 \includegraphics[trim = 15mm 0mm 15mm 0mm, clip, scale = 0.3, angle = 90]{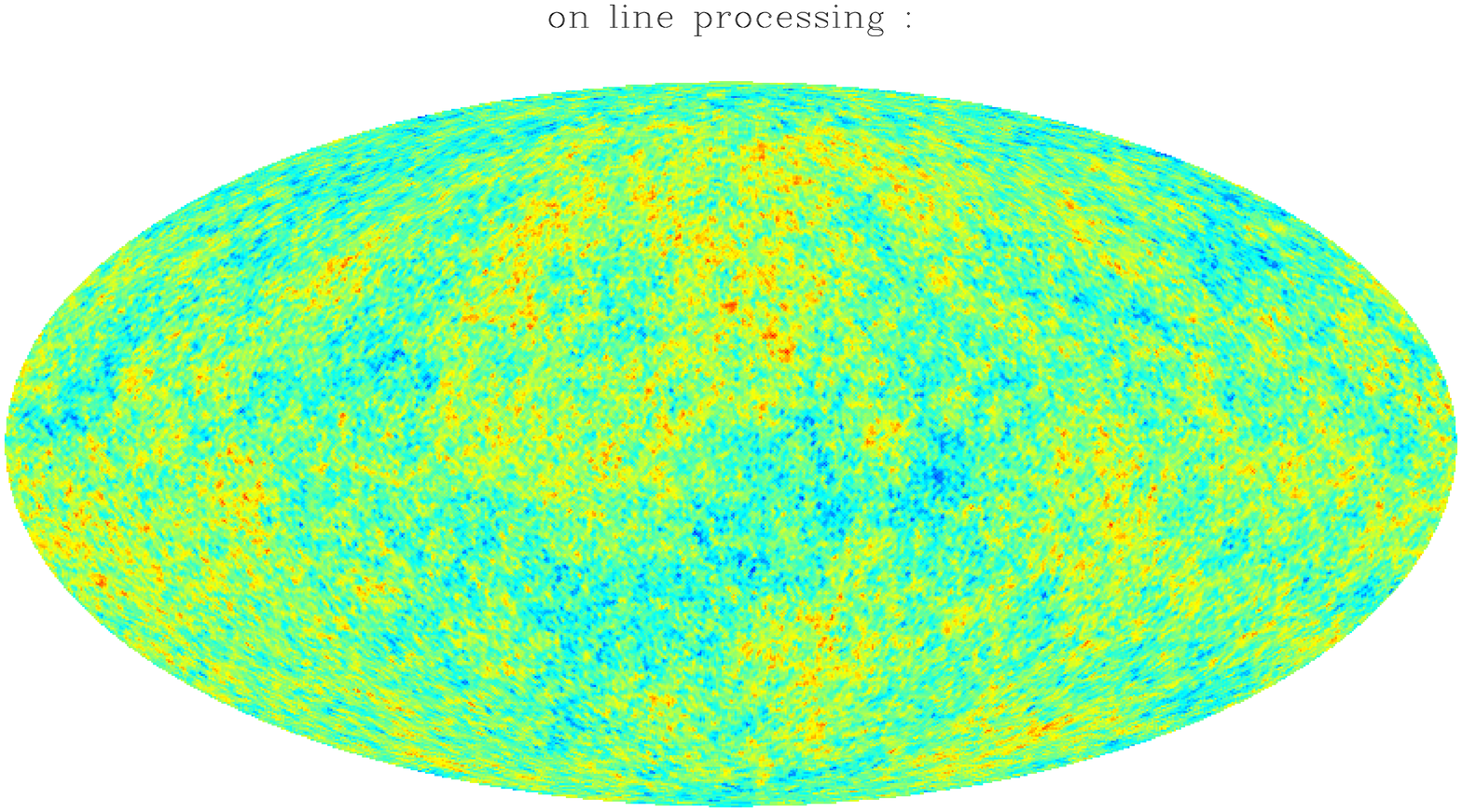}
 \includegraphics[trim = 0mm 0mm 15mm 0mm, clip, scale = 0.3, angle = 90]{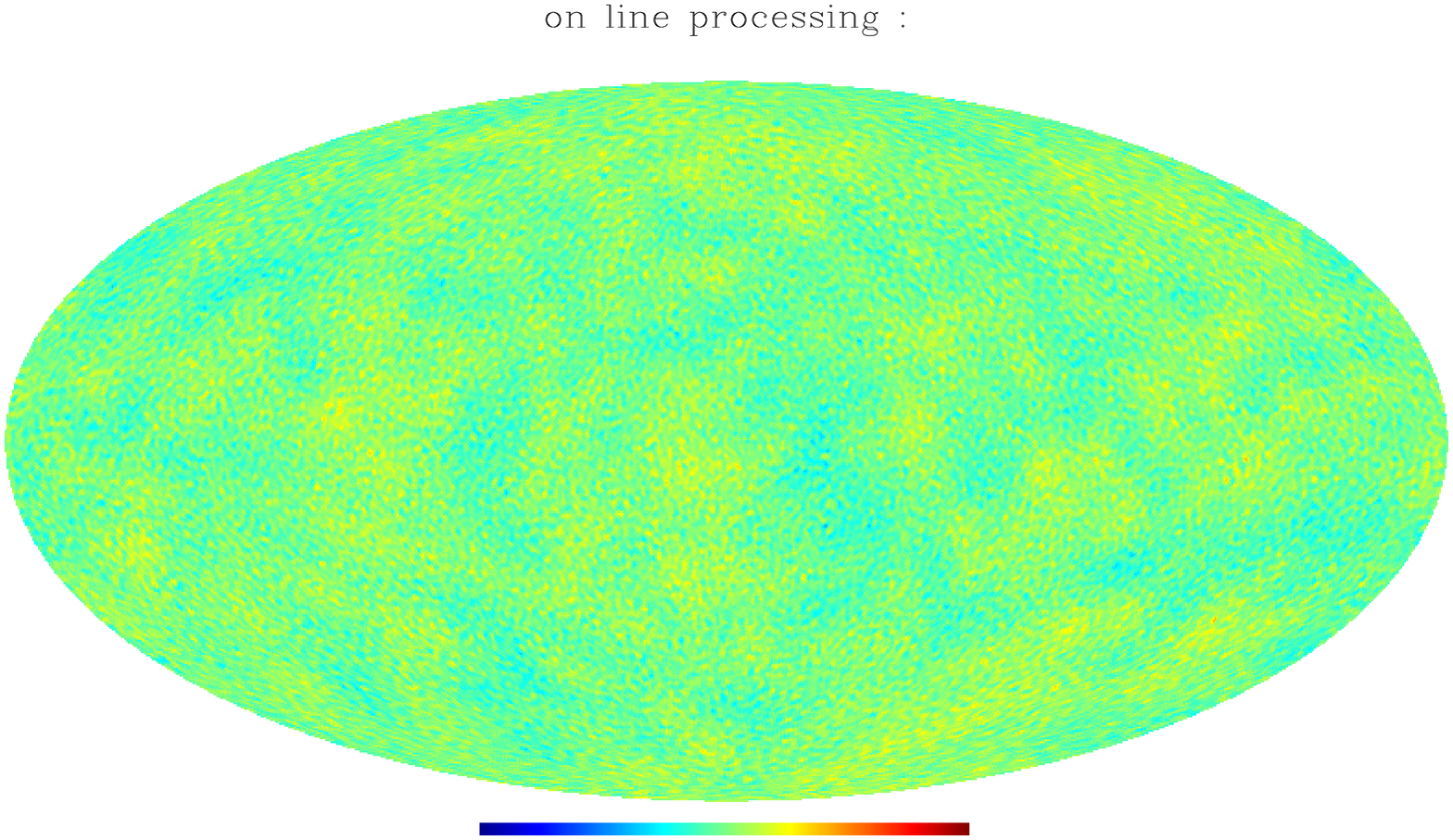}
 \caption{Realisation of the original CMB temperature map $\Tcmb$ (top panel),
   the reduced 
   temperature map $\Tred$ (middle panel) and the difference between the two
   for 
   comparison (bottom panel) in $\mu K$. We have chosen the same colour
 range from $-500 \mu K$ to $500 \mu K$ for all maps.}
 \label{maps}
\end{figure}

In Fig. \ref{maps}, we plot a realisation of the original temperature
map $\Tcmb$ (top panel), the reduced temperature map $\Tred$ (middle panel)
and the 
difference of the two, $\left(C_l^{\TerrObs,\Eobs}/C_l^\Eobs\right)
a_{lm}^\Eobs$, for comparison (bottom panel). The realisations 
were created using the HEALPix package \citep{healpix}.

\begin{figure}
 \centering
 \includegraphics{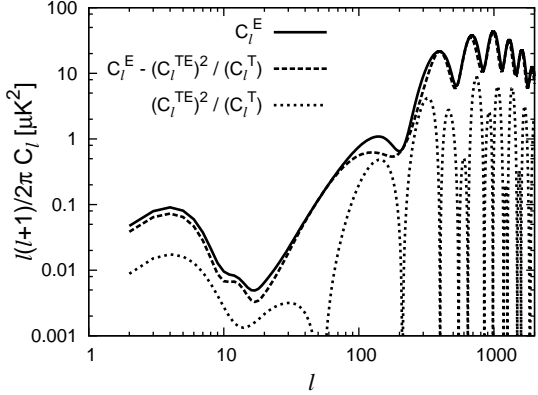}
 \caption{Reduction of the variance in the detection of secondary
   polarisation signals by using the information contained
   in temperature data.
   Shown are the CMB E-mode power spectrum $C_l^\Ecmb$
   (solid), and the template-free part of
   the reduced E-mode power spectrum $C_l^\Ecmb -
   \left(C_l^{\Tcmb,\Ecmb} \right)^2 / C_l^\Tcmb$ (dashed),
   together with the part of the CMB power spectrum coming from the
   'known' part of the E-mode fluctuations which we infer from
   the temperature map,
   $\left(C_l^{\Tcmb,\Ecmb} \right)^2 / C_l^\Tcmb$ (dotted).}
 \label{spectraPol}
\end{figure}

Note that all of what we have done works equally well for reducing
the E-mode polarisation map when trying to detect a secondary signal
contained in the polarisation data. One has to simply exchange the
roles of $T$ and $E$ in the derivation. This was partly already done
by \cite{jaffe}, who used the information contained in the CMB
temperature map for predicting a polarisation map from it. 
The equivalent plot to
Fig. \ref{spectra} for this scenario is given in Fig. \ref{spectraPol}.
The likelihood for the case of simultaneously detecting a temperature
template $\Ttempl$ and a polarisation template $\Etempl$ is derived in
Appendix \ref{prob_fact}.

In practice, the accuracy to which we can measure the E-map is limited
by galactic foregrounds $\Efg$, the most important of which are synchrotron
radiation and dust emission of the Milky Way. 
Uncertainty in the measured E-map makes the reduction
of the temperature power spectrum less efficient, because the power contained
in the foreground noise, $C_l^\Efg$, enhances the observed 
E-mode power spectrum $C_l^\Eobs \approx C_l^\Ecmb + C_l^\Efg + C_l^\Edet$.
The prediction of a
realistic signal-to-noise ratio for our method would require a
detailed study of foreground effects, detector noise, and scanning
strategies, which is beyond the scope of this work.

\section{Example: the ISW effect}\label{isw}

Let us now apply our method to the ISW effect. That is, our signal
template $\Ttempl$ is now an ISW template which we obtain from a Wiener
filter reconstruction of the LSS, which can be shown to be optimal for
the purpose of ISW detection \citep{mona}. 
We assume the best-case scenario of having perfect (noiseless)
LSS and CMB data. In other words, we neglect
the detector noise $\Tdet$ and $\Edet$, which is safe on the largest scales,
where cosmic variance dominates \citep{afshordi_manual}.
We furthermore neglect residual galactic foregrounds $\Tfg$ and $\Efg$
as well as the shot-noise in the observed galaxy
distribution, and assume that we have an ideal galaxy survey that covers the
whole sky and goes out to a redshift of at least two. Then our signal
template is exact, $\Ttempl = \Ts \equiv \Tisw$, and the residual
$(\Tcmb - \Tisw) 
\equiv 
\Tprim$ is simply given by the primordial CMB fluctuations, which are created
at the surface of last scattering (we have ignored other secondary
effects here). We further assume $\Tisw$ to be uncorrelated with the
primordial fluctuations $\Tprim$, which is a safe assumption because
they are created on very different scales \citep{x_ray_boughn}. We can
then write $C_l^{\Tcmb,\Ttempl} \equiv C_l^{\Tcmb,\Tisw} = C_l^\Tisw$.

\begin{figure}
 \centering
 \includegraphics{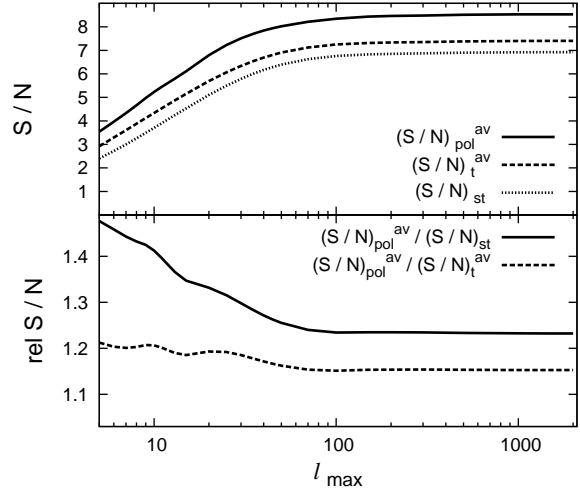}
 \caption{Comparison of the cumulative signal-to-noise ratios for
   $z_\Max =  2$. {\bf 
   Top panel:} Average signal-to-noise ratio of the optimal
   polarisation method $(S/N)_\pol^\av$
   (solid), of the optimal temperature-only
   method $(S/N)_t^\av$ (dashed), and 
   signal-to-noise ratio of the standard method $(S/N)_{\rm st}$ (dotted) versus the maximal
   multipole considered in the analysis. 
 {\bf Bottom panel:} Ratio
 of the signal-to-noise of the optimal polarisation method with the
 one of the
  standard method (solid) and with the one of the optimal
  temperature-only method 
 (dashed).}
 \label{compareSToN}
\end{figure}

The signal-to-noise ratio for the detection of the ISW signal,
eq. (\ref{StoNpol}), then reduces to
\be
\left(\frac{S}{N}\right)_\pol^2 =
\sum_l  \frac{ (2l+1)\, \widehat
    C_l^\Tisw}{ C_l^\Tprim 
    - \left(C_l^{\Tprim,\Ecmb}\right)^2 / C_l^\Ecmb }.
\ee
As we said before, the signal-to-noise ratio depends on the specific
LSS realisation in our Universe via $\widehat C_l^\Tisw$. We can infer its
probability distribution from the distribution of $\Tisw$ by using the
central limit theorem for the distribution of $\left(S/N\right)^2$ and
deriving the distribution for $S/N$ from that \citep[see
  also][]{mona}\footnote{This will provide accurate results for
  multipoles $l \gg 1$, 
however, is a coarse approximation in the regime $l \sim 1$.}.
We then average
the signal-to-noise ratio over this probability distribution in order
to compare it to the signal-to-noise ratio of the standard method
and the average signal-to-noise ratio of the optimal temperature-only method,
both described in \cite{mona}.
Recall that the signal-to-noise ratio one obtains for the standard
method is given by 
\be
\left(\frac{S}{N}\right)_{\rm st}^2 =
\sum_l  \frac{ (2l+1)\,
    C_l^\Tisw}{ C_l^\Tprim + C_l^\Tisw }.
\ee
The cumulative
signal-to-noise ratios versus the maximal multipole $l_\Max$ used in
the analysis are plotted in Fig. \ref{compareSToN}. Here we have
assumed the ideal galaxy survey described above. 
We see that including the polarisation data in the analysis increases
the signal-to-noise ratio by 16 per cent as compared to the optimal
temperature-only method, and by 23 per cent as compared to
the standard method. 
Note that we only included the linear ISW effect 
in Fig. \ref{compareSToN}. Beyond a multipole of about $l
\approx 100$, non-linear effects start to play a crucial role
\citep{cooray_rs}, which
could change the plot for $l > 100$. However, we see that for the
linear ISW effect, there is hardly any contribution for such high multipoles.

\begin{figure}
 \centering
 \includegraphics{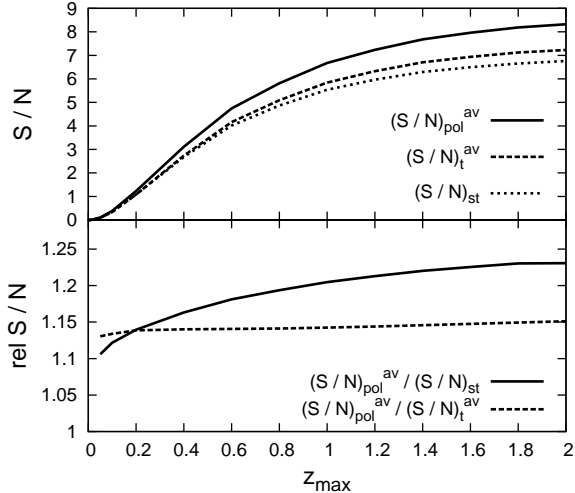}
 \caption{Comparison of the signal-to-noise ratios versus the maximal
   redshift $z_\Max$ of the galaxy survey. {\bf Top panel:} Average
   signal-to-noise ratio of the optimal polarisation
   method $(S/N)_\pol^\av$ (solid), of the optimal temperature-only
   method $(S/N)_t^\av$ (dashed) 
   and signal-to-noise ratio of the standard method $(S/N)_{\rm st}$
   (dotted). {\bf Bottom 
     panel:} Ratio
   of the signal-to-noise of the optimal polarisation method with the
   one of the
   standard method (solid) and with the one of the optimal
   temperature-only method 
   (dashed). We see that
   with polarisation data included, the signal-to-noise is significantly
   enhanced even for low redshifts.}
 \label{zPlot}
\end{figure}

Let us now look at the enhancement of the signal-to-noise ratio for
shallower LSS surveys. We use the same approximation as in
\cite{mona}, i.e. we introduce a sharp cut-off in redshift and
redefine everything beyond that redshift as primordial
fluctuations. This introduces a correlation between what we consider
the ISW and primordial fluctuations, which we would not have
if we had used a proper Wiener filter based template $\Ttempl$ for redefining
$\Tisw$. However, for getting a rough picture of the redshift
dependence, this approximation is good enough\footnote{The ratio of
  this neglected coupling to the template strength gets large for
  small $z_\Max$. Our estimates are therefore less accurate in this regime.}.  
We plot the
redshift-dependence of the signal-to-noise ratios of the three methods in
Fig. \ref{zPlot}. We also plot the ratio of the
signal-to-noise of the optimal polarisation method with the one
of the standard method (solid) and with the one of the
optimal temperature-only method (dashed).
Note that the enhancement of the signal-to-noise ratio
w.r.t. the optimal temperature-only method is almost constant in
redshift. This is quite clear from the fact that we have reduced the
{\it primordial} noise with the polarisation data, and neither the primordial
noise nor the reduction of the latter depend on redshift.
Therefore, the
reduction of the noise from including polarisation data is always the same,
independent of how deep in redshift our survey goes, and the
signal-to-noise ratio is already significantly enhanced for currently
available surveys.
For example, for a maximal redshift of $z_\Max
\approx 0.3$, which is the maximal redshift for the SDSS main galaxy
sample, we have a better signal-to-noise by about 16 per
cent as compared to the standard method. The additional enhancement for higher
redshifts of our 
signal-to-noise ratio w.r.t. the standard method comes from working
conditional on the galaxy data, as we have described in detail in
\cite{mona}.

\section{Conclusions} \label{conclusions}

The detection of secondary effects on the CMB remains a challenge,
because the amplitudes of these effects are much smaller than those of 
primordial CMB fluctuations. The techniques for detecting such
secondary signals are all based on the existing cross-correlation between 
the LSS and the signal in question.
However, in all of these studies, chance correlations of
primordial CMB fluctuations with the LSS
are the dominant source of noise in the analysis.

We have presented a way of reducing the noise coming from
primordial temperature fluctuations by simply subtracting the part of
the temperature map which is known from the polarisation data.
Effectively, only the unknown part of the temperature fluctuations
then contributes to the variance of the signal estimate. 

As presented here, our method can be generically applied to all secondary
effects. However, in this work we
have used a Gaussian approximation for the uncertainty in the signal
template, which may not be optimal for effects on smaller scales,
such as the RS effect, the kinetic SZ effect or lensing. We leave the
extension of our method to non-Gaussian noise models for future work.

We calculated the achievable reduction in primordial noise for perfect
(noiseless) data using the example of the ISW effect, and obtained a
signal-to-noise ratio of up to 8.5. This corresponds to an enhancement
of the signal-to-noise ratio by 16 per 
cent as compared to our optimal temperature-only method,
independent of the depth of the LSS survey.
In comparison to the standard method, the signal-to-noise ratio is
enhanced by 23 per cent for a full-sky galaxy survey which goes out to a
redshift of at least two.
When using the SDSS main galaxy sample, which has a maximal redshift of
about $z_\Max \approx 0.3$, our signal-to-noise
ratio is still enhanced by about 16 per cent as compared to the standard
method.

The variance reduction achieved with this method will significantly
improve the detection of all kinds of secondary effects on the CMB, where a
spatial template constructed from non-CMB data can be created.
This stresses the importance of accurate measurements of
primordial polarisation fluctuations even for non-primordial signal
detection and analysis.
The upcoming Planck Surveyor Mission, as well as more future
experiments like
PolarBeaR\footnote{\texttt{http://bolo.berkeley.edu/polarbear/index.html}}
or CMBPol\footnote{\cite{cmbpol}, \texttt{http://cmbpol.uchicago.edu}} 
will allow us to benefit from polarisation for the
detection of secondary CMB signals in the way presented in this work.

\section*{acknowledgments}

The authors would like to thank Martin Reinecke and Andr{\'e} Waelkens
for their extensive help with HEALPix. We would also like to thank Simon
D. M. White, Cheng 
Li and Thomas Riller for useful discussions and comments. We acknowledge the
use of the HEALPix package and CMBEASY.

\bibliographystyle{mn2e}
\bibliography{bibl.bib}

\begin{appendix}

\section{Proof of the factorization of the likelihood}\label{prob_fact} 

We now explicitely prove the factorization of the likelihood
in eq. (\ref{fact1}) into a reduced temperature part and a
polarisation part, as given in eq. (\ref{fact}). We will do this for
the more general case that we not only have a signal template
$\Ttempl$ for 
the temperature part, but also a non-zero template $\Etempl$ for the
polarisation part. 
In this case, the covariance matrix is
\be
\covL = \left( 
\begin{array}{cc}
C_l^\TerrObs & C_l^{\TerrObs,\ErrEobs} \\
C_l^{\TerrObs,\ErrEobs} & C_l^\ErrEobs  \\
\end{array} 
\right),
\ee
instead of the simplified one given in eq. (\ref{covdef}).
Here, $\ErrEobs$ is defined as $\ErrEobs \equiv \Eobs - \Etempl$.
The inverse of the covariance matrix is given by
\bea \nonumber
\covL^{-1} &=& \frac{1}{C_l^\TerrObs C_l^\ErrEobs -
  \left(C_l^{\TerrObs,\ErrEobs}\right)^2}\\
&&\times \left(  
\begin{array}{cc}
C_l^\ErrEobs & -C_l^{\TerrObs,\ErrEobs} \\
-C_l^{\TerrObs,\ErrEobs} & C_l^\TerrObs  \\
\end{array} 
\right).
\eea
We first rewrite the
exponent of $\g(a_{lm}^d - a_{lm}^\tau , \covL)$ in eq. (\ref{fact1})
by inserting the inverse of $\covL$:
\bea \nonumber
\!\!\!\!\! &&\!\!\!\!\! \left(a_{lm}^\Tobs\! - a_{lm}^\Ttempl,
  a_{lm}^\Eobs\! - a_{lm}^\Etempl\right) \covL^{-1} \!
  \left(a_{lm}^\Tobs \! - a_{lm}^\Ttempl,
  a_{lm}^\Eobs \! - a_{lm}^\Etempl\right)^\dagger \\ \nonumber
\!\!\!\!\! &=& \!\!\!\!\!
\left[
\left|a_{lm}^\TerrObs\right|^2 - 2
    \left(C_l^{\TerrObs,\ErrEobs}/C_l^\ErrEobs\right) 
    Re \left(a_{lm}^\ErrEobs a_{lm}^\TerrObs\right) \right. \\ \nonumber
\!\!\!\!\! && \!\!\!\!\! \left. + \left(C_l^\TerrObs/C_l^\ErrEobs\right)
    \left|a_{lm}^\ErrEobs\right|^2
\right]\\ \nonumber
\!\!\!\!\! && \!\!\!\!\! / \left[
C_l^\TerrObs - \left(C_l^{\TerrObs,\ErrEobs}\right)^2/C_l^\ErrEobs\right]
\\ \nonumber
\!\!\!\!\! &=& \!\!\!\!\! \frac{\left|a_{lm}^\TerrObs -  \left(C_l^{\TerrObs,\ErrEobs}/C_l^\ErrEobs\right)
    a_{lm}^\ErrEobs\right|^2}
{C_l^\TerrObs - \left(C_l^{\TerrObs,\ErrEobs}\right)^2/C_l^\ErrEobs} \\ \nonumber
\!\!\!\!\! && \!\!\!\!\! + \frac{\left|a_{lm}^\ErrEobs\right|^2}{C_l^\ErrEobs} \\
\!\!\!\!\! &\equiv& \!\!\!\!\! \frac{\left|a_{lm}^\Tred - a_{lm}^\Ttempl \right|^2}
{\covTredL} 
+ \frac{\left|a_{lm}^\Eobs - a_{lm}^\Etempl\right|^2}{C_l^\ErrEobs} 
\label{zerfall}
\eea
where we have completed the square in the second last step and used
a generalised definition of the reduced temperature map and power
spectrum, which 
we had introduced in eq. (\ref{red}), in the last step. Similarly, we can
decompose the determinant of $\covL$:
\bea \nonumber
\left|\covL\right| &=& C_l^\TerrObs C_l^\ErrEobs -
\left(C_l^{\TerrObs,\ErrEobs}\right)^2 \\ \nonumber
&\equiv& \covTredL C_l^\ErrEobs.
\label{determinante}
\eea
Inserting eqs (\ref{zerfall}) and (\ref{determinante}) into
$\g(a_{lm}^d - a_{lm}^\tau , \covL)$ 
allows us to write
\bea \nonumber
\g(a_{lm}^d - a_{lm}^\tau , \covL) &=& \g(a_{lm}^\Tred - a_{lm}^\Ttempl ,
\covTredL) \\
&\times& \g(a_{lm}^\Eobs - a_{lm}^\Etempl,C_l^\ErrEobs).
\eea
In the case of the polarisation template $\Etempl$ being zero, this
expression reduces to the one in eq. (\ref{fact}).

\end{appendix}

\label{lastpage}

\end{document}